\documentclass[conference]{IEEEtran}
\IEEEoverridecommandlockouts
\usepackage{amsmath,amssymb,amsfonts}
\usepackage{algcompatible}
\usepackage[ruled,linesnumbered]{algorithm2e}
\usepackage{color}
\usepackage{graphicx}
\usepackage{textcomp}
\usepackage[dvipsnames]{xcolor}
\usepackage{paralist}
\usepackage{subcaption}
\usepackage{multirow}
\usepackage{graphics}
\usepackage{url}
\usepackage{cancel}
\usepackage{makecell}
\def\BibTeX{{\rm B\kern-.05em{\sc i\kern-.025em b}\kern-.08em
    T\kern-.1667em\lower.7ex\hbox{E}\kern-.125emX}}
\usepackage[colorinlistoftodos,prependcaption,textsize=footnotesize]{todonotes}

\usepackage{mathtools}
\DeclarePairedDelimiter{\ceil}{\lceil}{\rceil}
\DeclarePairedDelimiter{\floor}{\lfloor}{\rfloor}

\usepackage[noadjust]{cite}

\let\oldnl\nl
\newcommand{\nonl}{\renewcommand{\nl}{\let\nl\oldnl}}

\usepackage{amsmath,bm}

\usepackage{xcolor,cancel}

\newcommand{\gpk}{\texttt{GPK}} 
\newcommand{\lpk}{\texttt{LPK}}
\newcommand{\ipk}{\texttt{IPK}} 

\begin{document}
\bstctlcite{IEEEexample:BSTcontrol}

\newif\iffinal
\finalfalse

\iffinal
  \newcommand\guidance[1]{}
  \newcommand\writer[1]{}
  \newcommand\kshitij[1]{}
  \newcommand\scott[1]{}
  \newcommand\matthew[1]{}
  \newcommand\jong[1]{}
  \newcommand\ian[1]{}
  \newcommand\todd[1]{}
  \newcommand\igor[1]{}
  \newcommand\dave[1]{}
\else
  \definecolor{amber}{rgb}{1.0, 0.49, 0.0}
  \definecolor{darkblue}{rgb}{0,0,0.5}
  \definecolor{crimson}{rgb}{0.83, 0.0, 0.25}
  \definecolor{darkgreen}{rgb}{0,0.5,0}
  \definecolor{purple}{rgb}{0.5,0,0.5}
  \definecolor{brown}{rgb}{0.65, 0.16, 0.16}
  \definecolor{caribbeangreen}{rgb}{0.0, 0.8, 0.6}
  \definecolor{carmine}{rgb}{0.59, 0.0, 0.09}
  \definecolor{champagne}{rgb}{0.97, 0.91, 0.81}
  \definecolor{classicrose}{rgb}{0.98, 0.8, 0.91}
  \definecolor{corn}{rgb}{0.98, 0.93, 0.36}
  \definecolor{cornflowerblue}{rgb}{0.39, 0.58, 0.93}
  \definecolor{darkelectricblue}{rgb}{0.33, 0.41, 0.47}
  \newcommand\guidance[1]{\todo[inline]{#1}}
  \newcommand\writer[1]{\color{red} -- #1}
  \newcommand\kshitij[1]{{\todo[inline,color=green]{Kshitij: #1}}}
  \newcommand\scott[1]{\todo[inline,color=red]{Scott: #1}}
  \newcommand\matthew[1]{{\todo[inline,color=champagne]{Matthew: #1}}}
  \newcommand\jong[1]{{\todo[inline,color=amber]{Jong: #1}}}
  \newcommand\ian[1]{{\todo[inline,color=corn]{Ian: #1}}}
  \newcommand\todd[1]{{\todo[inline,color=classicrose]{Todd: #1}}}
  \newcommand\igor[1]{{\todo[inline,color=caribbeangreen]{Igor: #1}}}
  \newcommand\dave[1]{{\todo[inline,color=orange]{Dave: #1}}}
  \newcommand\jieyang[1]{{\todo[inline,color=caribbeangreen]{Jieyang: #1}}}
\fi

\title{Accelerating Multigrid-based Hierarchical Scientific Data Refactoring on GPUs}

\author{\IEEEauthorblockN{Jieyang Chen,
Lipeng Wan, 
Xin Liang\IEEEauthorrefmark{1},
Ben Whitney, 
Qing Liu\IEEEauthorrefmark{2},
David Pugmire, 
Nicholas Thompson,\\
Jong Youl Choi,
Matthew Wolf, 
Todd Munson\IEEEauthorrefmark{3},
Ian Foster\IEEEauthorrefmark{3}\IEEEauthorrefmark{4},
Scott Klasky
}
\IEEEauthorblockA{
Oak Ridge National Laboratory, Oak Ridge, TN, USA
}
\IEEEauthorblockA{\IEEEauthorrefmark{1}
Missouri University of Science and Technology, Rolla, MO, USA
}
\IEEEauthorblockA{\IEEEauthorrefmark{2}
New Jersey Institute of Technology, Newark, NJ, USA
}
\IEEEauthorblockA{\IEEEauthorrefmark{3}
Argonne National Laboratory, Lemont, IL, USA
}

\IEEEauthorblockA{\IEEEauthorrefmark{4}
University of Chicago, Chicago, IL, USA
}

\{chenj3, wanl, whitneybe, pugmire, thompsonna, choij, wolfmd, klasky\}@ornl.gov\\
xlang@mst.edu qliu@njit.edu 
\{tmunson, foster\}@anl.gov
}

\maketitle

\linespread{0.97}
\selectfont


\begin{abstract}
Rapid growth in scientific data and a widening gap between computational speed and I/O bandwidth make it increasingly infeasible to store and share all data produced by scientific simulations.
Instead, we need methods for reducing data volumes: ideally, 
methods that can scale data volumes adaptively so as to enable negotiation of performance and fidelity tradeoffs in different situations.
Multigrid-based hierarchical data representations hold promise as a solution to this problem, 
allowing for flexible conversion between different fidelities
so that, for example, data can be created at high fidelity and then transferred or
stored at lower fidelity via logically simple and mathematically sound operations. 
However, the effective use of such representations has been hindered until now by the relatively high costs of creating, accessing, reducing, and otherwise operating on
such representations.
We describe here highly optimized data refactoring kernels for GPU accelerators that enable efficient creation and manipulation of data in multigrid-based hierarchical forms.
We demonstrate that our optimized design can achieve up to 250 TB/s aggregated data refactoring throughput---83\% of theoretical peak---on 1024 nodes of the Summit supercomputer.
We showcase our optimized design by applying it to a large-scale scientific visualization workflow and the MGARD lossy compression software.

\end{abstract}


\setlength{\textfloatsep}{0pt}


\section{Introduction}


With the dawn of the big data era, managing the massive volume of data generated by data-intensive applications becomes extremely challenging, particularly for scientific simulations~\cite{alexander2020exascale, Wan:2019} running on leadership-class high-performance computing (HPC) systems and experiments running on federated instruments and sensor platforms. 
For instance, the XGC dynamic fusion simulation code \cite{ku2009full,chang2004numerical} from  the Department of Energy (DOE)'s Princeton Plasma Physics Laboratory can generate 1 PB every 24 hours when running on the DOE's fastest supercomputers, and may soon generate 10 PB  per day.  
The Square Kilometer Array (SKA) \cite{taylor2004science} plans to generate data at 1~PB/s within 10--20 years. Few storage systems can keep up with such data rates.
Moreover, even if all data could be stored, the high costs of 
processing them with standard multi-pass analysis routines often lead to significant degradation in overall scientific productivity \cite{Wan:2017,Wan:hpcc:2017}.

There are no universal solutions to the many technical and domain-specific challenges of managing the overwhelming amount of complex and heterogeneous scientific data, and current approaches are usually passive and based on rules of thumb. 
For instance, scientists may decimate in time by reducing output data frequency by some arbitrary factor (e.g., writing one of every 1000 simulation steps). 
Although such approaches can effectively reduce the amount of data written to storage, they increase the risk of missing novel scientific discoveries, as discarded data may contain important features. 
Moreover, the limited capacity of fast storage such as parallel file systems means that data are eventually moved to slower storage, such as archival storage systems.
For example, on Oak Ridge National Laboratory (ORNL)'s Summit supercomputer, data can only be kept on the parallel file system for 90 days before it is either moved to archival storage systems such as HPSS \cite{hpss} or permanently deleted. 
Once moved to archival storage, it can take weeks or longer to retrieve for analysis. 

In plotting a course to address these data management challenges, it is important to remember the perspective of the end user, the domain scientist, for whom a dataset is valued not for its size in bytes but for the scientific information it contains.
A dataset need not be especially large to capture some feature of interest, and in fact the most valuable insights often come from just a small portion of the original data.
A domain scientist seeking to answer a question using a dataset would ideally be able to retrieve only the smallest subset or reduced representation of the data necessary to answer the question to the desired level of accuracy.
It is challenging to support this workflow with existing data compression techniques given the great variety of analyses a scientist might need to run, since the data would need to be compressed and stored separately for each analysis and each level of accuracy required, resulting in high computational and storage costs.

\emph{Data refactoring} is the capability of building a data representation in a hierarchical form such that a reader can easily, efficiently, and transparently access data at varying degrees of fidelity.
To enable this capability, new algorithms such as multigrid-based hierarchical data refactoring~\cite{ainsworth2018multilevel,ainsworth2019multilevel, ainsworth2019multilevel2} have recently been developed by the applied mathematics community.
That data refactoring approach
models a dataset with a series of hierarchically organized \emph{coefficient classes}, such that an approximation of the original data with a specified fidelity can be reconstructed by using different numbers of coefficient classes.
We call the process of building coefficient classes \emph{decomposition} and the process of reconstructing data from coefficient classes \emph{recomposition}.

\begin{figure*}[t]
    \centering
    \includegraphics[width=0.9\textwidth]{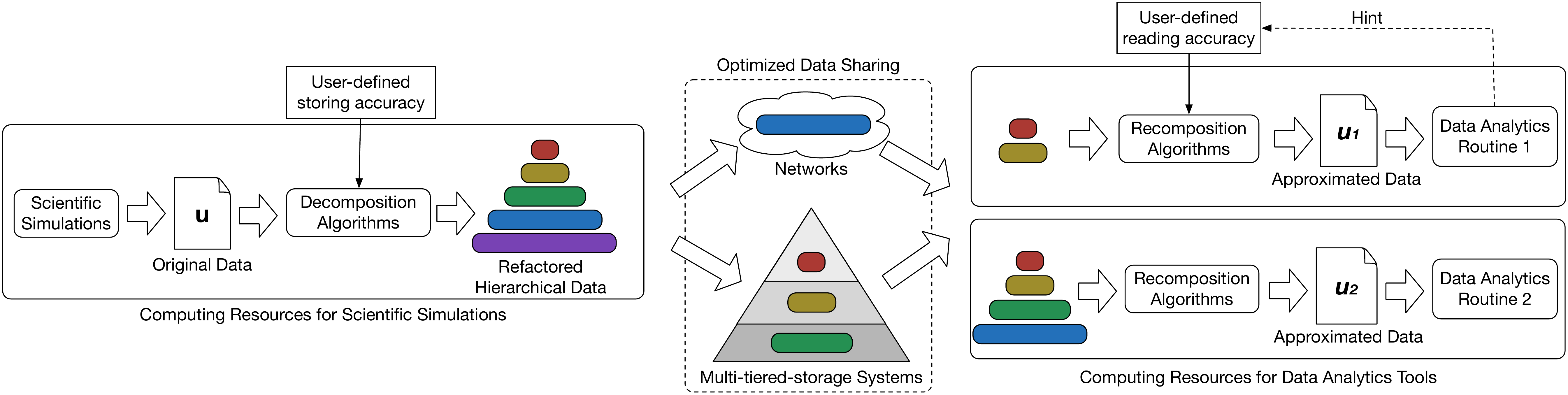}
    \caption{Example of hierarchical data refactoring helping optimize data movement in scientific workflows by intelligently moving each coefficient class across networks and/or multi-tiered-storage systems, based on available capacity and bandwidth.
    }
    \label{data-refac-exp}
    \vspace{-1em}
\end{figure*}


Hierarchical data refactoring gives both data producers (e.g., scientific simulations) and consumers (e.g., data analysis routines) the flexibility to store, transport, and access data to satisfy space and/or accuracy requirements.
For example, data sharing between two coupled scientific applications~\cite{CODAR2020} can be optimized by intelligently moving coefficient classes through multi-tiered-storage systems (e.g., storage systems containing non-volatile memory, magnetic disks, and tapes) \cite{Wan:2014, Wan:jpdc:2017} and/or networks based on available capacity and bandwidth.
In Figure~\ref{data-refac-exp}, simulation data are refactored into five coefficient classes and then shared with data analysis routines via multi-tiered-storage systems and networks.
When accuracy can be estimated based on the number of selected coefficient classes, users can control the accuracy of the reconstructed data while storing and reading the data.
If user-defined accuracy requirements indicate that information encoded in the first four coefficient classes are enough for subsequent data analyses, then the fifth coefficient class can be ignored.
Then, the four coefficient classes can be intelligently shared over the storage systems and network based on their size, available bandwidth/capacities, and accuracy requirements from data analysis routines.
In the figure, Data Analysis Routine~1 needs only two coefficient classes to achieve desired accuracy, while Routine~2 needs four.
The ability to choose a reduced number of coefficient classes allows users to reduce data movement costs substantially.


As great as the benefits of reduction in data movement and management costs may be, if the decomposition and recomposition routines are too expensive, then the total process is less useful in production.
The use of Graphics Processing Units (GPUs) for scientific computations that can be adopted to the streaming execution model has increased significantly due to the high parallel computational power and memory throughput of GPUs.
As the algorithms involved in multigrid-based hierarchical data refactoring are highly parallelizable, using GPUs to accelerate its routines is attractive.
Also, we anticipate that if used with merging GPU communication technologies~
\cite{li2018tartan, li2019evaluating} (e.g., NVLink, GPUDirect RDMA, etc.), GPU data refactoring would be greatly beneficial for speeding up data sharing for both CPU- and GPU-based scientific applications.

We focus here on accelerating the two major routines, decomposition and recomposition, in multigrid-based data refactoring on GPUs and evaluating the benefit for producer and consumer applications.
Although the multigrid-based algorithms are naturally parallelizable, achieving good performance requires carefully designed parallel algorithms together with deep optimizations for GPU architectures.
Our specific contributions, and the sections in which they are described, are as follows.

In \S\ref{sec:design}, 
we describe the first multigrid-based data refactoring routines for modern GPU architectures, 
and present 
systematic optimizations for multigrid-based data refactoring at three levels: instruction level, kernel level, and program-structure level. These optimizations can balance both minimizing memory footprint and improving memory access efficiency. 
    
In \S\ref{sec:eval}, 
we demonstrate our design by implementing the state-of-the-art non-uniform multi-dimensional multigrid-based data refactoring algorithms of Ainsworth et al.~\cite{ainsworth2018multilevel,ainsworth2019multilevel, ainsworth2019multilevel2},
and show that our methods perform well on both a consumer-class desktop and the Summit supercomputer, achieving 145$\times$ and 14$\times$ speedups compared with state-of-the-art CPUs and GPUs, and 250 TB/s throughput on 1024 Summit nodes.

In \S\ref{sec:showcase},
we use two common scenarios in scientific computing to showcase our work: 1) reducing data movement costs between simulations and in situ visualization applications; and 2) speeding up lossy compression for scientific data. 
\section{Background}\label{sec:background}
\subsection{Theory of multigrid-based hierarchical data refactoring}
The multigrid-based hierarchical data refactoring developed by Ainsworth et al. supports nonuniformly-spaced structured multidimensional data, commonly found in scientific computations, by using
hierarchical representations to approximate data. Specifically, they decompose data from fine grid representation to coarse grid representation in an iterative fashion, with a global correction to account for the impact of missing grid nodes in each iteration.
If we use functions to represent the discrete values continuously, the decomposition from fine grid level $l$ to coarser grid level $l-1$ can be formulated with the notation in Table~\ref{tab:notation} as follows,

\vspace{-1em}
\begin{equation}
\label{e1}
\underbrace{Q_{l-1}u}_{\substack{\text{Projection} \\ \text{onto} V_{l-1}}} = \underbrace{Q_{l}u}_{\substack{\text{Projection} \\ \text{onto} V_{l}} }- \underbrace{(I-\Pi_{l-1})Q_{l}u}_{\text{Coefficients}} +  \underbrace{(Q_{l-1}u - \Pi_{l-1}Q_{l}u)}_{\text{Corrections}}
\end{equation}

\noindent
where the piecewise linear function $u$ takes the same values as the original data for each node; $Q_{l-1}u$ and $Q_lu$ are the function approximations of $u$ at levels $l-1$ and $l$, respectively; $(I-\Pi_{l-1})Q_{l}u$ is the difference between the values of the fine grid nodes at level $l$ and their corresponding piecewise linear approximations; and $(Q_{l-1}u - \Pi_{l-1}Q_lu)$ is the global correction.
According to Eq.~\eqref{e1}, two major steps are involved at each level of the multigrid decomposition: 1) compute coefficients for the current multigrid level $l$; and 2) compute the global correction and add it to the nodes in the next coarse grid (level $l-1$). In what follows, we introduce how to compute coefficients and corrections.

\begin{table}[t]
\centering
\caption{Notation used in algorithms, formulations, figures}
\label{tab:notation}
\begin{tabular}{|c|l|}
\hline
Symbol &  Description\\ \hline
$u$ &  Function represented by the original data.\\ \hline
$N_l$ &  Nodes at grid level $l$.\\ \hline
$C_l$ &  Coefficients at grid level $l$.\\ \hline
$V_l$ &  Function space with respect to $N_l$.\\ \hline
$Q_l$ &  The $L^2$ projection onto $V_l$.\\ \hline
$\Pi_l$ &  The piecewise linear interpolant in space $V_l$.\\ \hline
$a \rightarrow b$ & $b$ is calculated using $a$.\\ \hline
\end{tabular}
\end{table}

\subsubsection{Compute coefficients}
\begin{figure}[t!]
    \centering
    \begin{subfigure}[t]{0.24\textwidth}
    \includegraphics[width=\textwidth, trim=4mm 4mm 1mm 3mm, clip]{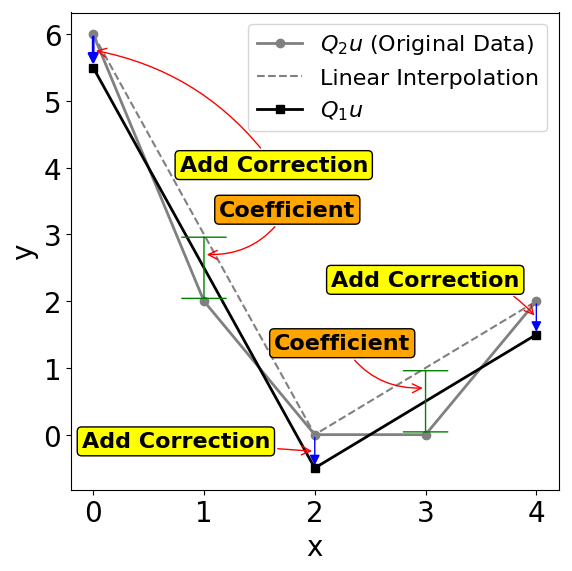}
    \vspace{-1em}
    \caption{Decomposition at $l=2$}
    \end{subfigure}
    \begin{subfigure}[t]{0.24\textwidth}
    \includegraphics[width=\textwidth, trim=2mm 4mm 3mm 3mm, clip]{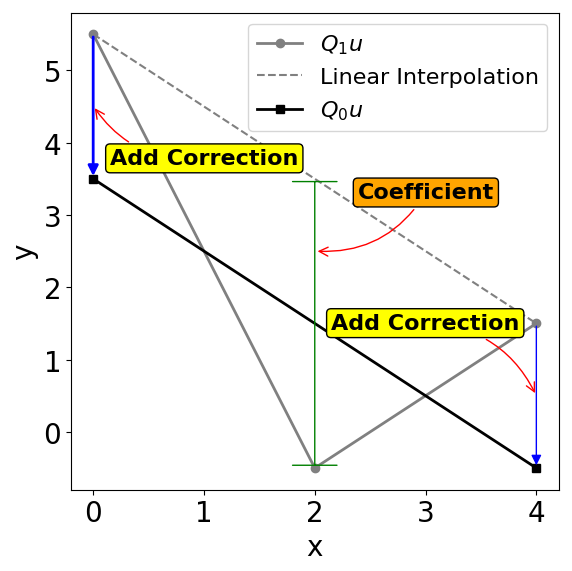}
    \vspace{-1em}
    \caption{Decomposition at $l=1$}
    \end{subfigure}
    \caption{Example of decomposing a 1D dataset produced from discretizing a quadratic function: $y = x^2-5x+6$}
    \label{example-1d}
\end{figure}

The coefficients store the difference between the data approximated by nodes at levels $l$ (i.e., $N_l$) and $l-1$ (i.e., $N_{l-1}$) before corrections are added.
Since $N_{l-1}$ is contained in $N_l$, its nodes have the same values in both levels; thus the nonzero differences only occur on nodes in $N_l \setminus N_{l-1}$. 
Figure~\ref{example-1d} shows how coefficients are calculated along one dimension through linear interpolation.
It can be generalized to multi-dimensional cases easily by using multi-linear interpolations for approximation.


\begin{figure*}[ht]
    \centering
    \includegraphics[width=0.9\textwidth]{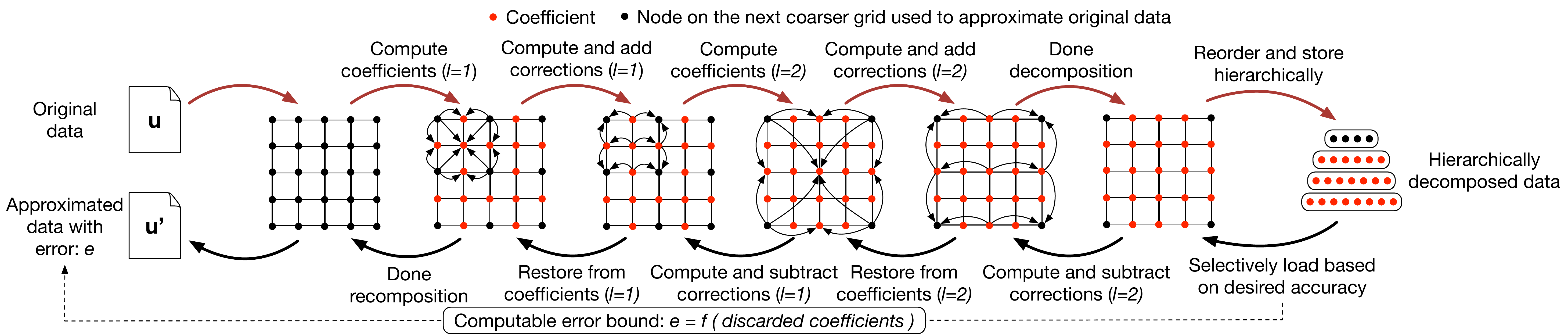}
    \caption{Multigrid-based data refactoring: Decomposition (left to right); recomposition (right to left).}
    \label{refac-recomp-example}
    \vspace{-1em}
\end{figure*}

\subsubsection{Compute correction}
Ainsworth et al.\ prove that the correction is the orthogonal projection of the calculated coefficients at grid level $l$ onto $V_{l-1}$~\cite{ainsworth2018multilevel}; thus, adding the correction to the next coarse grid better approximates data in the current grid. 
To explain, we first define $z_{l-1}$ as the correction for grid at level $l-1$. From Eq.~\eqref{e1}, we have that:

\vspace{-1em}

\begin{equation}
\label{e2}
z_{l-1} - \underbrace{(I-\Pi_{l-1})Q_{l}u}_{\text{Coefficients}} = -(Q_{l} - Q_{l-1})u \in V_{l-1}^{\bot}
\end{equation}

\vspace{-0.5em}

If we apply $L^2$ projection at grid level $l-1$ (i.e., $Q_{l-1}$) to both sides of Eq.~\ref{e2}, it leads to a zero function since it belongs to $V_{l-1}^{\bot}$. 
Also, since $z_{l-1}$ is in $V_{l-1}$, $Q_{l-1}z_{l-1} = z_{l-1}$.
So, we can see that $z_{l-1}$ is the orthogonal projection of the coefficients onto $V_{l-1}$. Namely, $Q_{l-1}(I-\Pi_{l-1})Q_{l}u = z_{l-1}$.

The correction can thus be computed by solving a variational problem: find $z_{l-1} \in V_{l-1}$ such that $(z_{l-1}, v_{l-1}) = ((I-\Pi_{l-1})Q_{l}u, v_{l-1})$ for all $v_{l-1} \in V_{l-1}$.
Then, $z_{l-1}$ can be found by solving linear systems
$M_{l-1}z_{l-1} = f_{l-1}$
where $M_{l-1}$ is a tensor product of the mass matrices~\cite{mass} of each dimension, i.e., $ M_{l-1} = M^1_{l-1} \otimes M^2_{l-1} \cdots \otimes M^d_{l-1}$, where $d$ is the number of dimensions and $f_{l-1}$ is the load vector, which can be calculated using:
$f_{l-1} =  R_lM_l \text{vec}(C_l)$,
where $R_l$ is a transfer matrix that coverts basis functions from $V_l$ to $V_{l-1}$ and $C_l$ is the coefficient matrix at level $l$, which consists of computed coefficients at $N_l \setminus N_{l-1}$ and zeros at $N_{l-1}$.

\subsubsection*{Overall decomposition/recomposition process}
Figure~\ref{refac-recomp-example} illustrates this process 
on a 5$\times$5 2D dataset. 
The original data is on the left, and the refactored representation is on the right.
The decomposition process moves from left to right (i.e., from finest to coarsest grid) and involves four steps: computing coefficient and computing correction (II.A.1 and II.A.2) for each of the two levels.
For multi-dimensional data, the computation of correction is done by working on each dimension in a prescribed order~\cite{liang2020optimizing}; in this 2D example, it proceeds first along the rows and then along the columns.
Recomposition moves from right to the left: i.e., from coarsest to finest grid.
There are again four total stages, but these occur in the reverse order.
The approximation of the original data is produced after recomposition.
Based on how coefficients are omitted in recomposition, an error bound on data approximation can be computed~\cite{ainsworth2019multilevel}.

\subsection{Existing GPU-based data refactoring} 
The state-of-the-art MGARD~\cite{mgard} GPU-based data refactoring system
redesigns original serial algorithms to expose high parallelism to suit the many-core architecture of modern GPUs.
It achieves $O(n^3)$ thread concurrency for computing coefficients and $O(n^2)$ thread concurrency for computing corrections, and applies node reordering such that each kernel can take advantage of coalesced memory accesses. 
Theoretically, with large inputs, these levels of thread concurrency are more than enough to fully occupy GPU cores that can help achieve high data refactoring throughput. 
However, performance evaluation shows that it still suffers from underutilized memory throughput, achieving less than 10\% of theoretical peak.

\section{Designing GPU-accelerated data refactoring}
\label{sec:design}
We next discuss the design of our GPU-accelerated multigrid-based hierarchical data refactoring method. 
We first focus on the optimizations for each computing kernel involved in data refactoring.
We classify the computing patterns into three categories and propose three general kernel designs for GPUs.
Following the efficient kernel designs, we discuss optimizations to help each of the kernels efficiently work together so that their performance can be maximized.
Finally, we discuss design details about how to use heuristic auto tuning to maximize the refactoring throughput.

\subsection{Designing optimized GPU multigrid kernels}
Decomposition and recomposition each involve three major 
steps: 1) computing coefficients; 2) mass-transfer matrix multiplication; and 3) correction solver.
Based on their computation pattern, we can classify them into three categories: \textit{grid processing style}; \textit{linear processing style};  and \textit{iterative processing style}. 
We design kernels dedicated for each processing style. 

\subsubsection{Grid processing kernel (\gpk)}
Grid processing style has the characteristic of processing data in a grid-wise fashion.
Namely, it processes nodes within the domain of a grid in a certain resolution level (e.g., $N_l$) or between neighboring levels (e.g., $N_l$ and $N_{l-1}$).
In the multigrid-based data refactoring, the calculation of coefficients follows the grid processing style.
The major calculation is to compute the interpolation at nodes in $N_l \setminus N_{l-1}$ using nodal values in $N_{l - 1}$. 
Parallelization can favor either interpolation operations (i.e., parallelism $\propto O(N_l \setminus N_{l-1})$) or accessing nodal values (i.e., parallelism $\propto O(N_{l})$).  The former can lead to a less thread divergence, while the latter can achieve a higher memory access efficiency.
The computation of coefficients is a memory bound operation, as its time complexity is $O(n)$.
Therefore, it is essential to optimize in favor of memory access efficiency instead of computation.
This is also chosen in the state-of-the-art GPU data refactoring~\cite{mgard}.

\begin{figure}[ht]
    \centering
    \includegraphics[width=0.7\columnwidth]{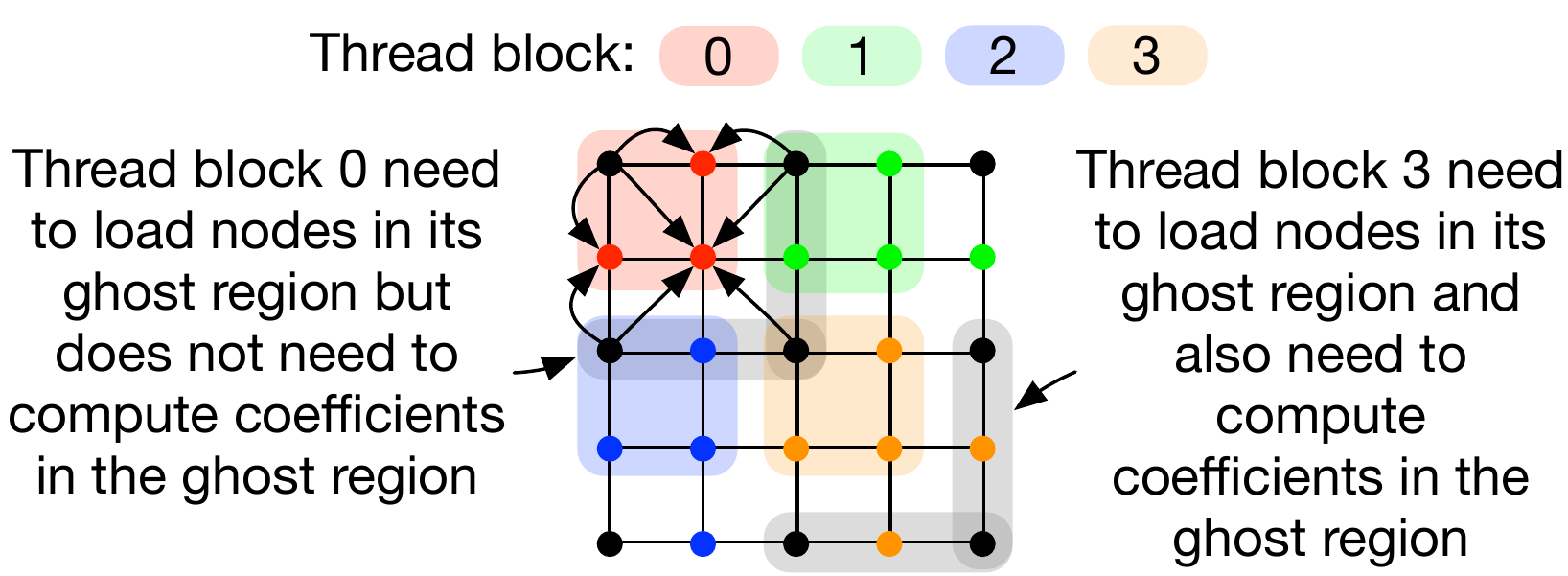}
    \caption{The workload of computing coefficients is distributed among 4 thread blocks. Calculating coefficients in the corresponding ghost regions is needed for some thread blocks (e.g., thread blocks 1, 2, and 3). }
    \label{gpk-ghost}
    \vspace{-1em}
\end{figure}

The key strategy they used to optimize for memory access is to use shared memory to cache a block of data for processing, of which the nodes values are loaded/stored in a coalesced-friendly fashion.
However, we identify that keeping efficient data movement on memory bound computations is not enough to achieve good performance. 
The level of thread divergence in a computation can still make a great impact on the overall performance and sometimes it can wrongly convert computation from memory bound to compute bound.
The reason is threefold: 1) high degrees of thread divergence can great increase the total cycle cost in computation; 2) variable floating point operation counts caused by different interpolation types further brings workload imbalance which leads to longer idling cycles; 3) as shown in Figure~\ref{gpk-ghost}, some thread blocks also need to calculate coefficients in the ghost region, which exacerbate the effect of thread divergence.

\begin{figure}[ht]
    \centering
    \includegraphics[width=0.9\columnwidth,trim=2.5mm 2mm 0mm 1mm,clip]{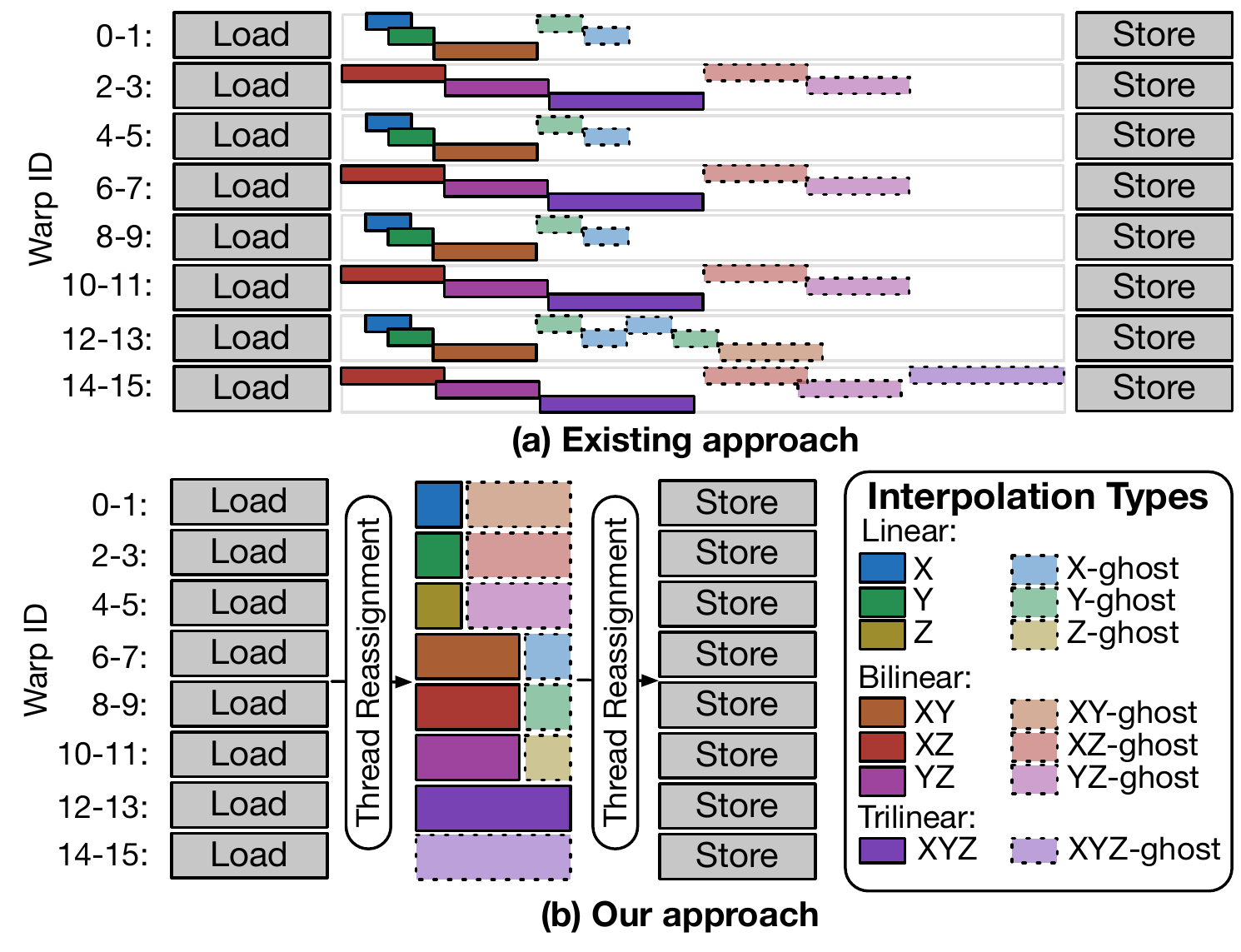}
    \caption{Conceptual flow of a thread block with $8\times8\times8$ threads (16 warps) calculating coefficients using the existing and our grid processing kernel \gpk. The thread reassignment strategy allows \gpk\ to greatly reduce thread divergence. }
    \label{gpk}
\end{figure}

However, we found that keeping efficient memory access patterns is not exclusive with having low thread divergence.
In designing our \gpk, we propose to decouple memory access and computation on nodal values in terms of thread-node assignment through a thread reassignment strategy.
Specifically, we use two different thread-node assignments for loading/storing nodal values and computing interpolations such that we maintain efficient coalesced memory access pattern while having one warp process the same type of interpolations along the same dimension.

Figure~\ref{gpk} shows the conceptual execution flow of one thread block using the existing approach and our proposed \gpk\ when computing coefficients.
As nodes in $N_{l}$ need to be shared with neighbors during interpolation operations,  we let each thread block coordinate work on a block of data and use shared memory as a scratch space. 
We organize threads such that threads in the same warp load values that are consecutive in memory to achieve efficient coalesced memory access patterns.
For computing, we apply a thread re-assignment strategy to achieve divergence-free execution.
Algorithm \ref{alg-coeff} shows how we calculate the thread-interpolation operation assignments that minimize thread divergence.
It is easy to see that the reassignment processing brings negligible computing overhead.


\SetKwInOut{KwInOut}{In/Out}
\SetKwInOut{KwIn}{In}
\SetKwInOut{KwOut}{Out}
\begin{algorithm}[t!]
\small
\caption{Thread re-assignment strategy}
\label{alg-coeff}
\SetKwFunction{FMain}{InterpolationType}
\SetKwProg{Fn}{Function}{:}{\KwRet}
\Fn{\FMain{}}{
$B_x, B_y, B_z \leftarrow$ Thread block size\\
$x, y, z \leftarrow$ Thread local indexes within thread block\\
$lane\_id, warp\_id \leftarrow x, y, z$ \\
$T\leftarrow$ 8 //total num.\ of warp group (4 for 2D)\\
$group\_id \leftarrow warp\_id/(B_x\times B_y\times B_z)/T$ \\
switch($group\_id$) {\\
\quad case 0: Linear-x(main) and Bilinear-xy(ghost)\\
\quad case 1: Linear-y(main) and Bilinear-xz(ghost)\\
\quad case 2: Linear-z(main) and Bilinear-yz(ghost)\\
\quad case 3: Bilinear-xy(main) and Linear-x(ghost)\\
\quad case 4: Bilinear-xx(main) and Linear-y(ghost)\\
\quad case 5: Bilinear-yz(main) and Linear-z(ghost)\\
\quad case 6: Trilinear-xyz(main)\\
\quad case 7: Trilinear-xyz(ghost)\\
}
}
\end{algorithm}

\subsubsection{Linear processing kernel (\lpk)} 
\begin{figure}[ht]
    \centering
    \includegraphics[width=0.5\textwidth]{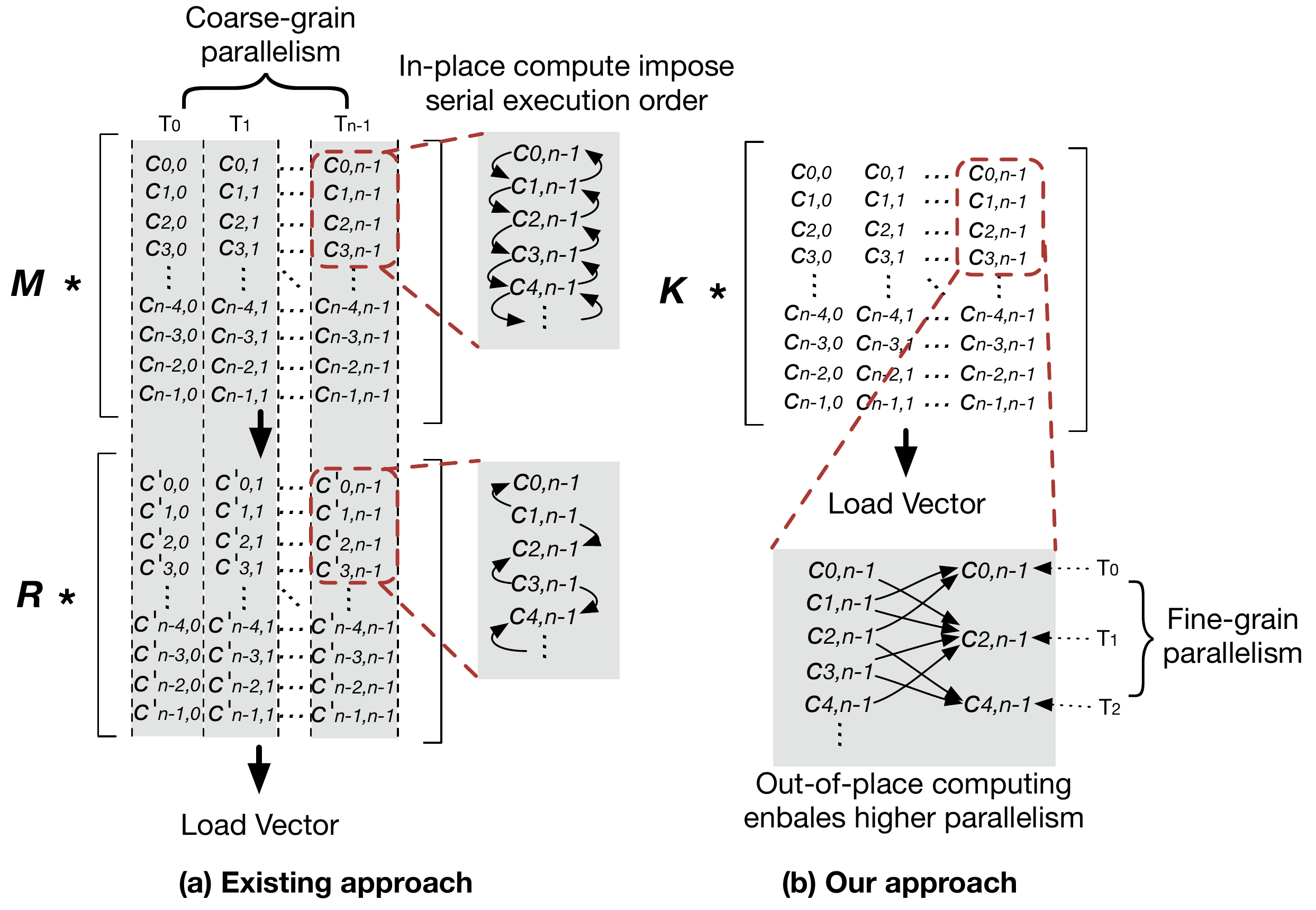}
    \caption{The conceptual workflow of mass and transfer multiplication using existing approach and proposed approach. Through optimizations, our approach achieves finer grain parallelism.}
    \label{lpk}
\end{figure}

The linear processing style computes stencil operations on elements in vectors along one dimension in a grid.
In multigrid-based data refactoring, when multiplying the mass and transfer matrices with computed coefficients, the computations become stencil operations, as the matrices are defined as:
\[
M_{ij}=\left\{\begin{matrix}
2(h_{i}+h_{i+1}) & \text{if~} i=j\\ 
h_{i} & \text{if~} |i-j|=1\\ 
0 & \text{else} 
\end{matrix}\right.
\]
\[
R_{ij}=\left\{\begin{matrix}
1 & \text{if~} i=j/2\\ 
r_{j-1} & \text{if~} i=(j-1)/2\\ 
1-r_{j} & \text{if~} i=(j+1)/2 \\
0 & \text{else} 
\end{matrix}\right.
\]
where $h_{i}$ is the spacing between the $i^{\textnormal{th}}$ node and the ${i+1}^{\textnormal{th}}$ node and $r_i = h_{i}/(h_{i}+h_{i+1})$.
As shown in Figure~\ref{lpk}(a), each value of each node needs to be computed using the original values of its neighbors, which means it cannot update its stored value unless all neighbors have finishing using its original value for computation.
Such data dependencies present a dilemma for kernel design: common out-of-place designs (i.e., element-wise parallelism) bring high parallelism but also high memory footprint; on the other hand, in-place design (i.e., vector-wise parallelism), used in \cite{mgard}, sacrifices the opportunity to exploit intrinsic parallelism.

To eliminate this dilemma, we design a novel linear processing kernel (\lpk) with four optimizations.
First, we change the original computation from in-place to out-of-place to achieve finer-grain parallelism.
Second, we merge the mass and transfer matrices to reduce computational costs.
We call the new matrix \texttt{mass-trans}, which is defined as:
$$K_{ij}=\left\{\begin{matrix}
(2+r_{j-2})h_{j-1}+(1+r_{j}) & \text{if~} i=j/2\\ 
(2r_{j-2}+1)h_{j-1}+2r_{j-2}h_{j-2} & \text{if~} i=(j-1)/2\\ 
(3-2r_{j})h_{j+1}+2r_{j+1}h_{j+1} & \text{if~} i=(j+1)/2 \\
r_{j-2}h_{j-2} & \text{if~} i=(j-2)/2\\ 
(1-r_{j})h_{j+1} & \text{if~} i=(j+2)/2\\ 
0 & \text{else} 
\end{matrix}\right.$$
Third, we use shared memory to cache a tile of nodes to allow sharing of coefficients (input) between different threads, so as to reduce total accesses to global memory.
Finally, to reduce extra memory footprint we use a kernel fusion technique to fuse the operation of copying coefficients with the multiplication of the mass-trans matrix with coefficients along the first dimension.
By eliminating the need to store a copy of the computed coefficient in the workspace, we avoid a large increase in the overall memory footprint.

\subsubsection{Iterative processing kernel (\ipk)}
The iterative processing style has the characteristic of processing nodes in a grid that contains strong data dependencies such that nodes have to be processed iteratively in a certain order. 
In multigrid-based data refactoring, the correction solver needs to solve for the corrections.
We use the Thomas algorithm~\cite{atkinson1985elementary}, which needs a forward and a backward pass on the load vector. 
Since the load vectors along one dimension can be solved independently, they can be solved in parallel.
This level of parallelization is well exploited in \cite{mgard}.
Specifically, they assign each thread to handle the solving process of one load vector independently.
Although this brings high thread concurrency with divergence free execution, it actually suffers from inefficient memory accesses for two reasons: first when solving vectors on leading dimension full coalesced memory access cannot be achieved (actual achieve efficiencies are about only 12\% and 25\% for single and double precision data); second, compared with \gpk\ and \lpk, \ipk\ only has $O(n^2)$ degrees of thread concurrency, which may bring less on-the-fly memory accesses to fully utilize the memory bandwidth. 

\begin{figure}[ht]
    \centering
    \includegraphics[width=\columnwidth,trim=2mm 0mm 2mm 2mm,clip]{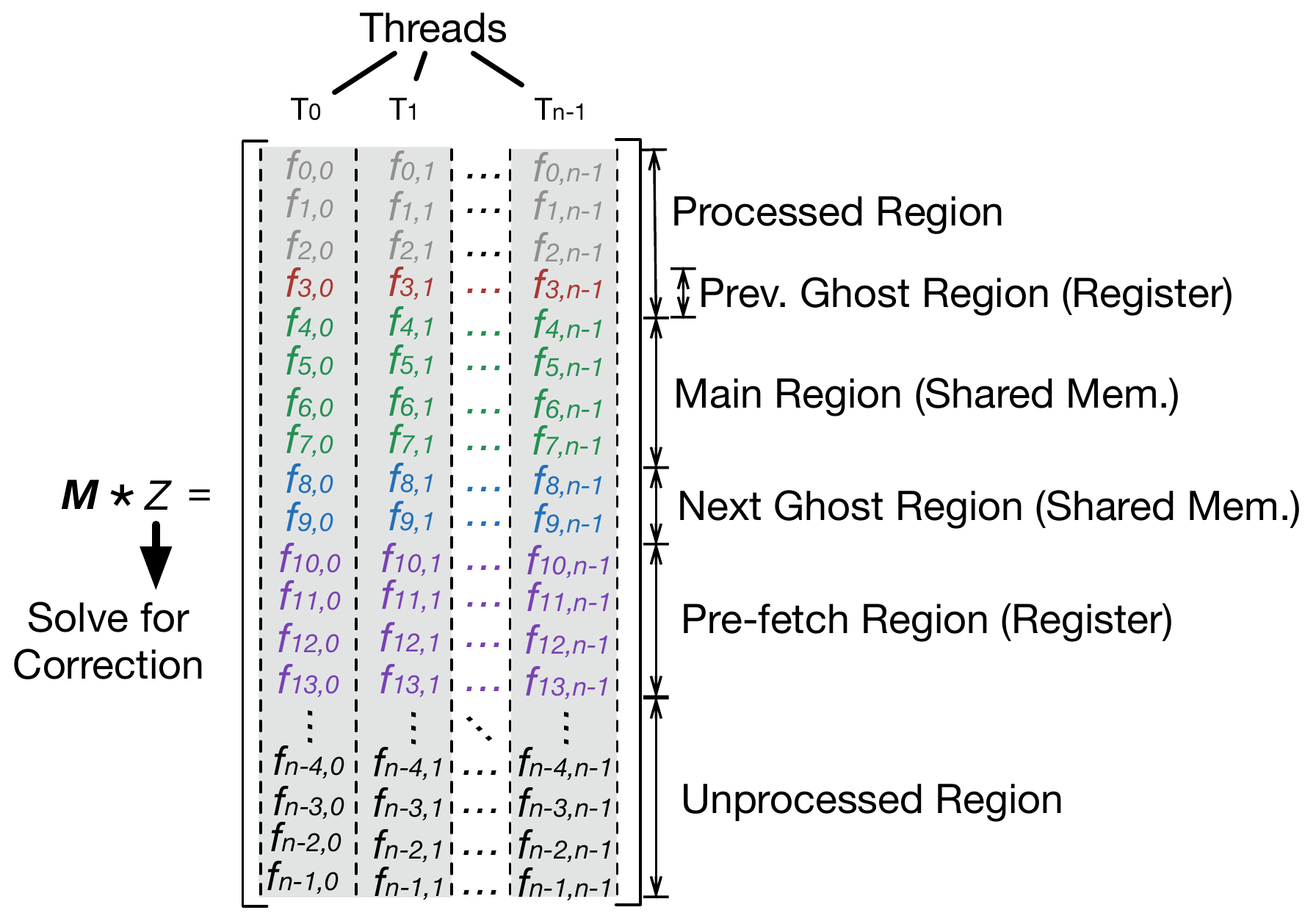}
    \caption{Correction solver designed following iterative processing kernel (\ipk). The node vectors are partitioned into six regions during processing. The use of shared memory ensures efficient coalesced memory accesses regardless of which dimension it is processing. Data prefetching further increases concurrency on memory accesses.}
    \label{mass-exp}
    \vspace{-1em}
\end{figure}

To address this issue, we proposed a novel processing kernel, \ipk, that can guarantee efficient coalesced memory access patterns with high concurrent memory accesses.
We first parallelize the vectors by assigning a batch to a thread block.
Since the update of each node depends on its neighboring elements, we use shared memory as scratch space to avoid polluting the un-processed nodes.
Specifically, we let each thread block iteratively work on a segment of load vectors at a time until the whole vector is updated.
Thus, as shown in Figure~\ref{mass-exp}, during the computation we divide the elements in the vectors into six regions: 1) the processed region stores updated elements (gray); 2) the main region consists of elements that the current iteration is working on (green); 3) and 4) due to dependence on the neighboring elements, the original values of elements in the two ghost regions (red and cyan) are needed to update the elements in the main region; 5) for better streaming processor utilization, we pre-fetch data needed for the next iteration (purple); and 6) we mark the unprocessed region as in block.
The regions move forward as the computation proceeds.
One challenge in designing the algorithm is to simultaneously consider maximizing coalesced global memory access patterns, minimizing bank conflict in accessing shared memory, and minimizing thread divergence.
We use a dynamic data-thread assignment strategy~\cite{chen2019tsm2,rivera2020tsm2x, chen2016online, chen2018fault, chen2016gpu} to optimize both the accessing and computation of coefficients.

\begin{figure}[ht]
    \centering
    \includegraphics[width=0.6\columnwidth]{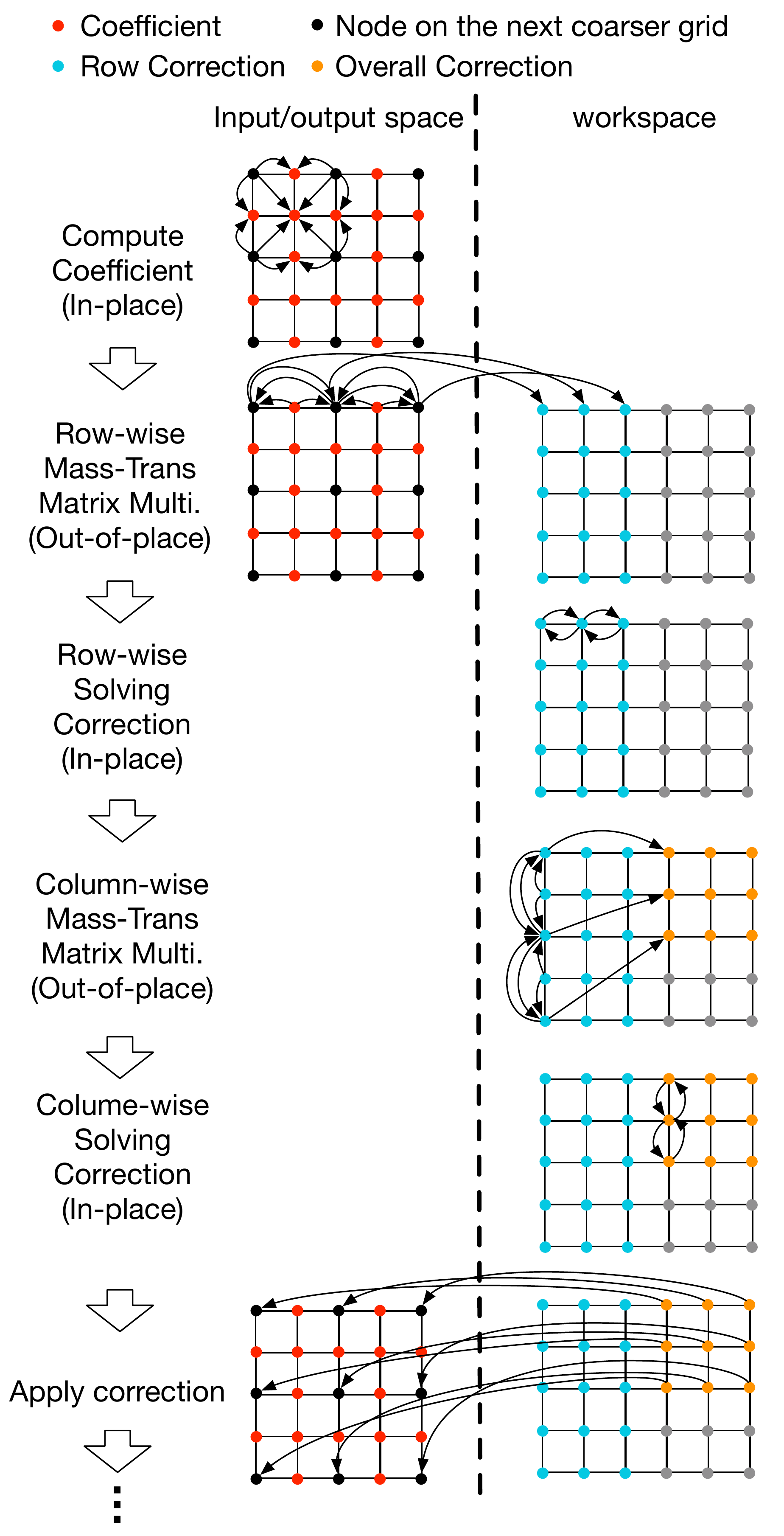}
    \caption{Overall decomposition process with optimizations.}
    \label{fig:overall}
\end{figure}

\subsection{Overall algorithms}

Figure \ref{fig:overall} shows how we use our optimized kernels to build data refactoring routines for multi-dimensional data on GPUs. 
For each level, the computed coefficients are also used for correction calculations.
This process involves altering the values of coefficients.
So, to preserve the values of previously computed coefficients, the correction is computed in a workspace.
In the state-of-the-art design~\cite{mgard}, the computed coefficients are first copied to the workspace before they are used for computing corrections, which limit the design's capability to do out-of-place computing unless using extra memory space.
Our optimization merges the copy of coefficients, with the first mass-trans matrix multiplication, so that it enables out-of-place computation.
We further extend out-of-place mass-trans matrix multiplication for processing other dimensios, which improves parallelism with a slight increase in memory footprint for the workspace.
In the state-of-the-art design, the workspace is of size $m\times n \times k$ and in our design its size is $(m+1)\times (n+1) \times k$, where $m$, $n$, and $k$ are the three dimensions of the input data.
The recomposition process is the opposite so we omit showing its process due to the page limit.

\subsection{Heuristic Performance Auto Tuning}
When launching each proposed kernel, choosing the execution parameters is important for achieving good performance, since even with the optimized design the parameters can still greatly impact the efficiency of memory accesses, warp divergence, context switch overhead, etc.
Auto tuning is an effective approach for searching the optimum configurations.
However, brute force search can be expensive and thus impractical.
Thus, we propose to use a heuristic auto tuning approach guided by theoretical performance models for our GPU data refactoring.
We first build performance models for the three kernels we proposed.
Among all tunable execution parameters, we find that the size of the thread block ($B_x, B_y, B_z$) plays an important role in determining each kernel's performance.
Since we eliminate the majority of the thread divergence and inefficient computations, we assume the memory load/store takes the majority time, so we only consider the total amount of memory transactions with their efficiency.
The estimated execution time of each kernel can be modeled as:
\begin{align*}
T_{\textrm{GPK}} = & \ceil{B_x+1/(S/L)} \cdot (S/L) \cdot (B_y+1) \cdot (B_z+1)\ \cdot \\
&\floor{N/B_x} \cdot \floor{N/B_y} \cdot \floor{N/B_z} \cdot 2 L\ \cdot \\
& (1/\text{Peak Mem. Band.})
\end{align*}
\begin{align*}
T_{\textrm{LPK}} = & \left(\ceil{B_x/(S/L)} \cdot S/L + 2 S/L \right) \cdot B_y \cdot B_z\ \cdot \\
& \floor{N/B_x} \cdot \floor{N/B_y} \cdot \floor{N/B_z} \cdot 2 L\ \cdot \\
& (1/\text{Peak Mem. Band.})
\end{align*}
\begin{align*}
T_{\textrm{IPK}} = & \left(\ceil{G/(S/L)} \cdot S/L + \ceil{B_x/(S/L)} \cdot S/L \cdot \ceil{N/B_x}\right)\  \cdot\\
& B_y \cdot B_z \cdot \floor{N/B_y}\ \cdot \floor{N/B_z} \cdot 2 L\  \cdot \\
&  (1/\text{Peak Mem. Band.})
\end{align*}
where $S$ is the number of bytes per memory transaction,
(32 in our test GPU); 
$L$ is bytes per float (4 for single, 8 for double); and
$G$ is the dimension of the next ghost region, set to $S/L$ so that ghost data can fit into exactly one memory transaction and do not consume too much shared memory. 
Table~\ref{auto-tuning} shows the ranking of estimated performance using seven typical thread block size configurations. 
Numbers in red represent the actual best configuration as determined by profiling.
We can see our performance model can help up predict relationship between different configuration in terms of performance with relative high accuracy.
It helps us narrow down the searching space for auto tuning.
For instance, in our following evaluation, we only let the auto tuning search and pick among the estimated top three configurations to save time.

\begin{table}[]
\centering
\caption{Ranking of estimated performance of seven typical thread block size configurations; actual best in red. }
\label{auto-tuning}
\begin{tabular}{|c|c|c|c|c|c|}
\hline
$B_z$ & $B_y$ & $B_x$   & GPK & LPK & IPK \\ \hline
2 & 2 & 2   & 7   & 7   & 7   \\ \hline
4 & 4 & 4   & 6   & 6   & 1   \\ \hline
4 & 4 & 8   & 4   & 5   & 2   \\ \hline
4 & 4 & 16  & \textcolor{red}{\textbf{2}}   & 4   & \textcolor{red}{\textbf{3}}   \\ \hline
4 & 4 & 32  & 1   & 3   & 4   \\ \hline
2 & 2 & 64  & 5   & 2   & 5   \\ \hline
2 & 2 & 128 & 3   & \textcolor{red}{\textbf{1}}   & 6   \\ \hline
\end{tabular}
\end{table}
\section{Experimental Evaluation}\label{sec:eval}
We evaluate our work on two GPU-enabled platforms.
Each node of the \textbf{Summit} supercomputer at ORNL is equipped with 6 NVIDIA \textbf{Volta} GV100 GPUs 
with 16 GB memory on each GPU and two 22-core (of which 21 cores/socket are accessible for computation) IBM POWER9 CPUs with 512 GB memory. 
\textbf{Turing} is a GPU-accelerated desktop with an NVIDIA RTX 2080 Ti GPU with 11 GB of memory and one 8-core Intel i7-9700K CPU with 32 GB of memory.

\begin{figure*}[ht!]
    \centering
    \begin{subfigure}[t]{0.32\textwidth}
    \includegraphics[width=\textwidth,trim=7mm 10mm 6mm 7mm, clip]{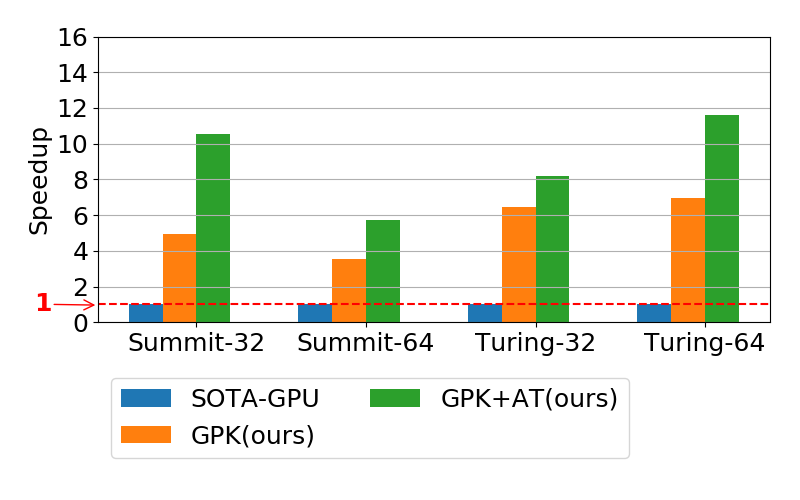}
    \caption{Coefficient calculation}
    \end{subfigure}
    \begin{subfigure}[t]{0.32\textwidth}
    \includegraphics[width=\textwidth,trim=7mm 10mm 6mm 7mm, clip]{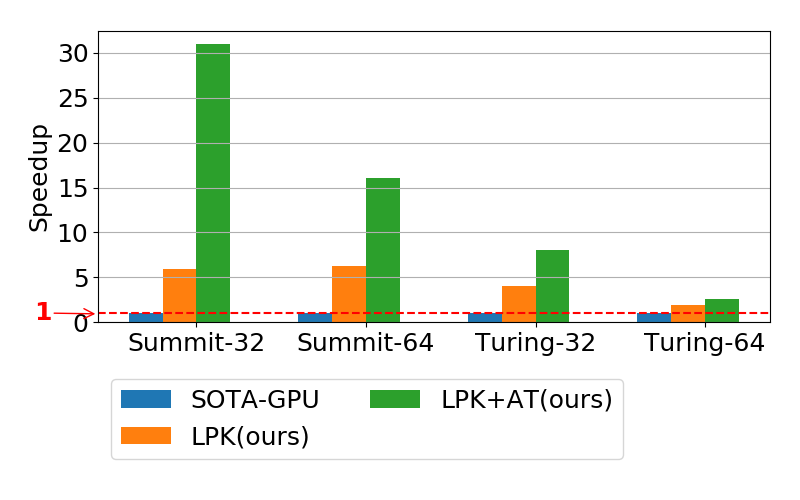}
    \caption{Mass-transfer matrix multiplication}
    \end{subfigure}
    \begin{subfigure}[t]{0.32\textwidth}
    \includegraphics[width=\textwidth,trim=7mm 10mm 6mm 7mm, clip]{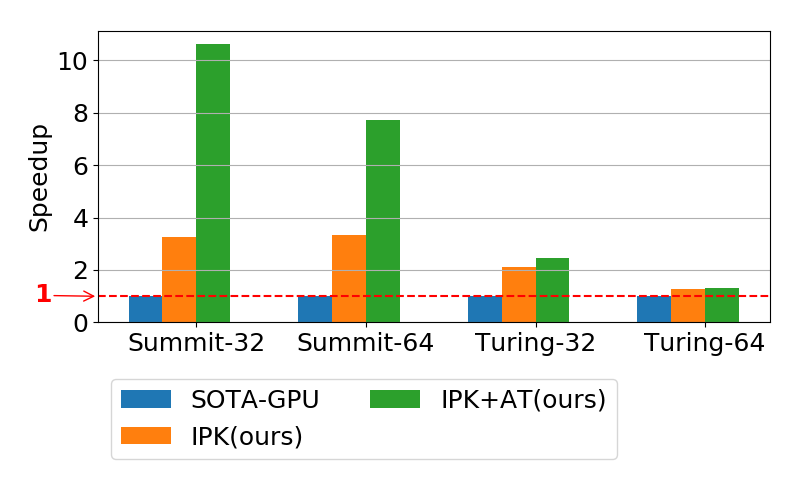}
    \caption{Correction solver}
    \end{subfigure}
    \caption{Speedups achieved through using our proposed processing kernels compared with the state-of-the-art GPU designs. 32 and 64 represent single and double precision input.}
    \label{kernel-speedup}
    \vspace{-1em}
\end{figure*}

\begin{figure*}[h]
    \centering
    \begin{subfigure}[t]{0.49\textwidth}
    \includegraphics[width=\textwidth]{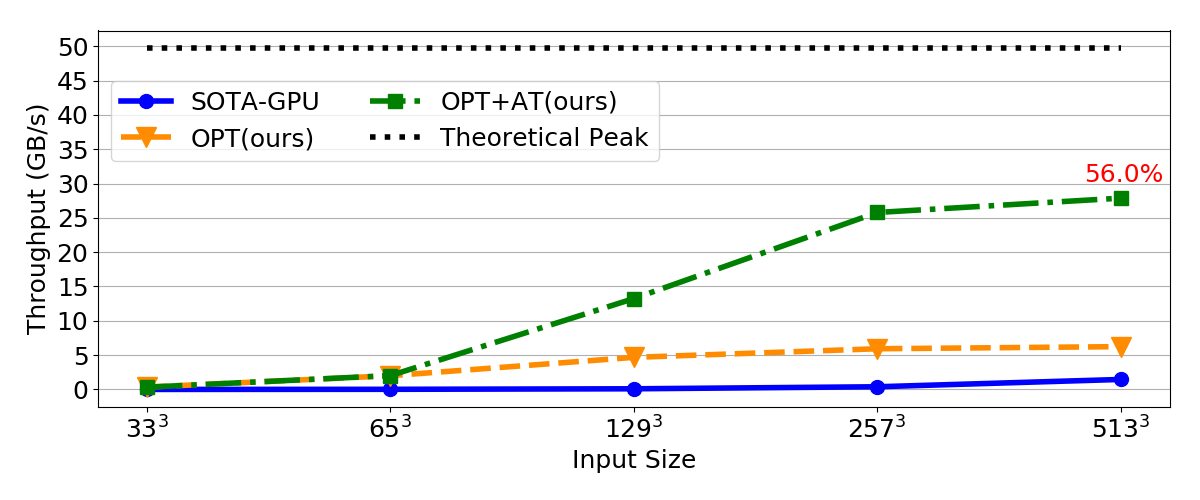}
    \caption{Volta (32-bit)}
    \end{subfigure}
    \begin{subfigure}[t]{0.49\textwidth}
    \includegraphics[width=\textwidth]{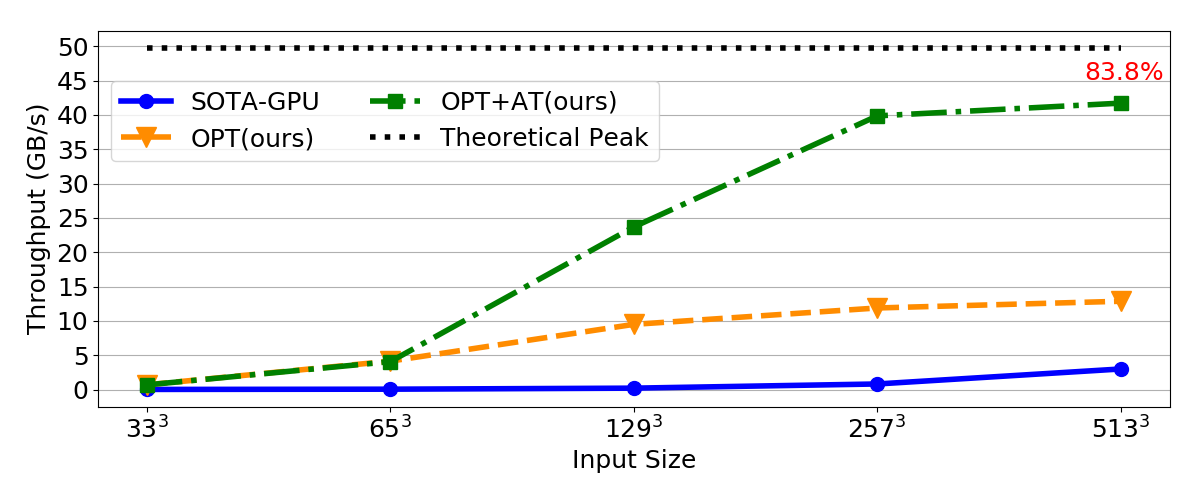}
    \caption{Volta (64-bit)}
    \end{subfigure}
    \begin{subfigure}[t]{0.49\textwidth}
    \includegraphics[width=\textwidth]{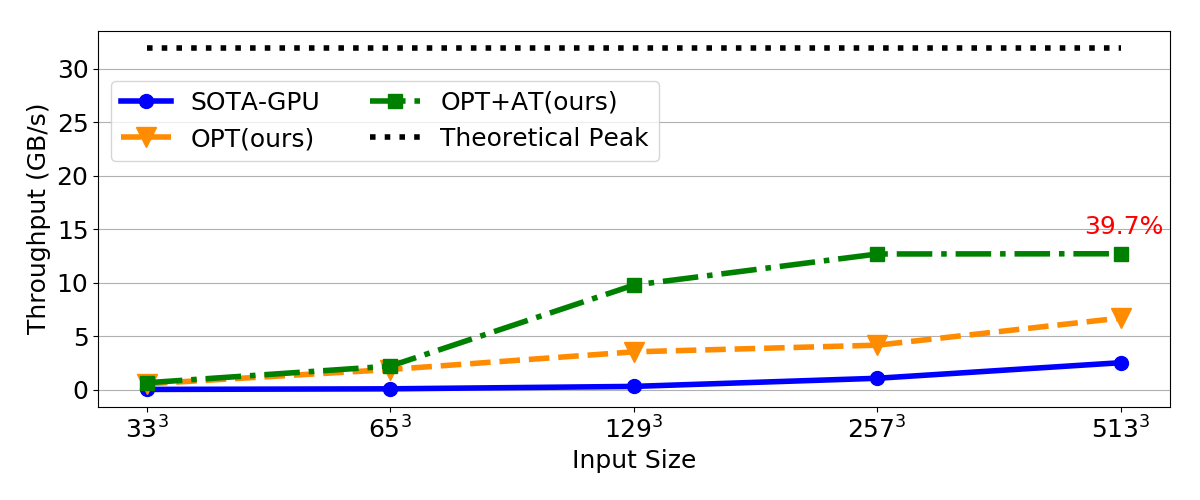}
    \caption{Turing (32-bit)}
    \end{subfigure}
    \begin{subfigure}[t]{0.49\textwidth}
    \includegraphics[width=\textwidth]{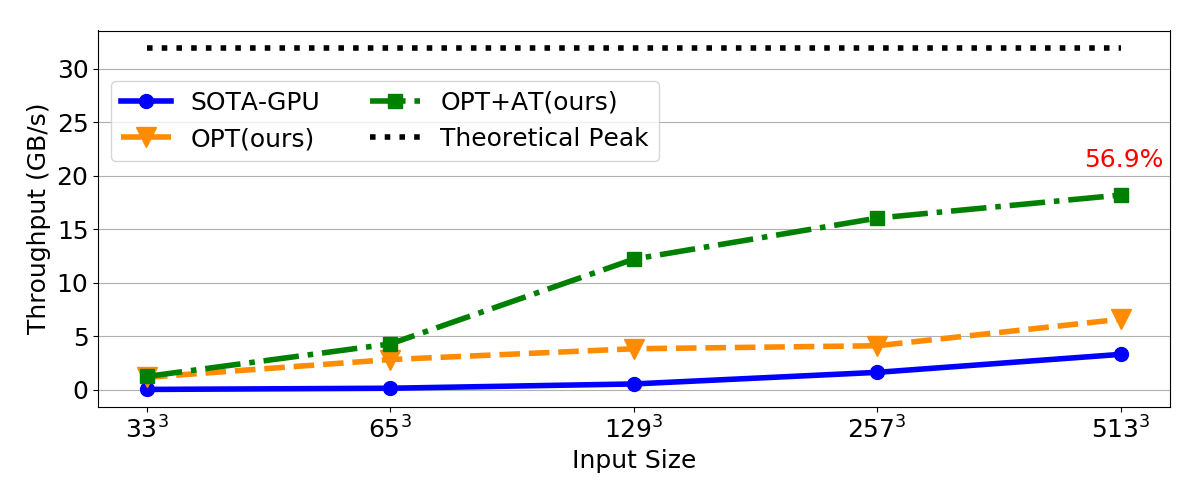}
    \caption{Turing (64-bit)}
    \end{subfigure}
    \caption{Data refactoring throughput on a single GPU}
    \label{fig:overall2}
    \vspace{-1em}
\end{figure*}

\subsection{Evaluation methodology}
We use datasets from a Gray--Scott reaction--diffusion simulation \cite{pearson1993complex, gscode}.
Each node in the input grid data is represented as single or double precision floating point values.
Note that our data refactoring algorithms have deterministic computation time complexity regardless of the values in the chosen dataset, so it will yield the same performance for any dataset with the same dimensions and size.
For simplicity, we let each dimension have the same size in our experiments.

We evaluate five different data refactoring implementations.
\begin{itemize}
    \item \textbf{SOTA-GPU}: We use the state-of-the-art GPU data refactoring in the MGARD lossy compression software~\cite{mgard} as our GPU baseline. Its design includes two performance tuning parameters: thread block size and number of CUDA streams. In our evaluation, we use the best performance achieved by hand tuning those parameters.
    \item \textbf{SOTA-CPU}: We use the state-of-the-art CPU data refactoring implemented in the MGARD lossy compression software~\cite{mgard}, parallelized with MPI for a fair comparison, as our CPU baseline.
    \item \textbf{OPT}: Our GPU data refactoring, which uses our novel grid/linear/iterative processing kernels (i.e., \gpk,~\lpk, and~\ipk) but not auto tuning.
    \item \textbf{OPT+AT}: OPT plus auto tuning.
\end{itemize}

\subsection{Evaluation on kernels}

We first show the performance improvement we achieve from accelerating the three major operations in data refactoring on GPUs.
Figure~\ref{kernel-speedup} shows speedups achieved on the three operations on the two GPU platforms with both single and double precision inputs. 
The input size is 513$\times$513$\times$513.
For single precision input, with the thread-level load-compute decoupled design, coefficient calculation with \gpk{} outperforms the existing design by 4.9$\times$ and 6.9$\times$ on Summit and Turing GPUs, respectively.
For mass-transfer matrix multiplication, with higher thread concurrency and data dependency free calculation, \lpk{} achieves 6.3$\times$ and 4.1$\times$ speedups on Summit and Turing GPUs, respectively. 
For correction solver, \ipk{} triples the performance on Summit and doubles the performance on Turing with the same level of thread concurrency as the state-of-the-art design, thanks to the more efficient memory access patterns.
Also, leveraging our heuristic auto tuning capability, the optimum configurations can be selected automatically, yielding additional 1.2--4.9$\times$ speedups compared with choosing one configuration for all kernels and input sizes.



\subsection{Evaluation on data refactoring on a single GPU}
Figure~\ref{fig:overall2} shows the end-to-end data refactoring throughput achieved on a single GPU with different input sizes.
(As decomposition and recomposition are symmetric processes, they have identical performance.)
To see how close the achieved data refactoring throughput is to the theoretical peak throughput, we estimate the theoretical peak by dividing the achievable single pass throughput
with the accumulated number of passes on the entire input data over the data refactoring process. 
(The achievable single pass throughput is the maximum throughput achievable when data are read and stored on GPU memory once. We measured it through a specially designed benchmark kernel that simultaneously reads and writes the same amount of data from and to the GPU memory without computation.)

The accumulated number of passes is calculated by summing the number of passes for all decomposition levels: \textit{passes per level} $\times \frac{1}{1-\frac{1}{8}}$. \textit{passes per level} = 1(coefficient calculation) + 1(copy to workspace) + 5.25(correction calculation) + 0.125(apply correction). 

The theoretical peaks for Summit and Turing GPUs
are 49.8 GB/s and 32.0 GB/s, respectively, for both single and double precision data. 
The state-of-the-art GPU data refactoring methods that we use as our baseline achieve only up to 10.4\% of the theoretical peak throughput;
our optimized GPU data refactoring achieves up to 83.8\% of theoretical peak.

\subsection{Evaluation on multi-node performance at scale}
\begin{figure}[ht]
    \centering
     \includegraphics[width=\columnwidth]{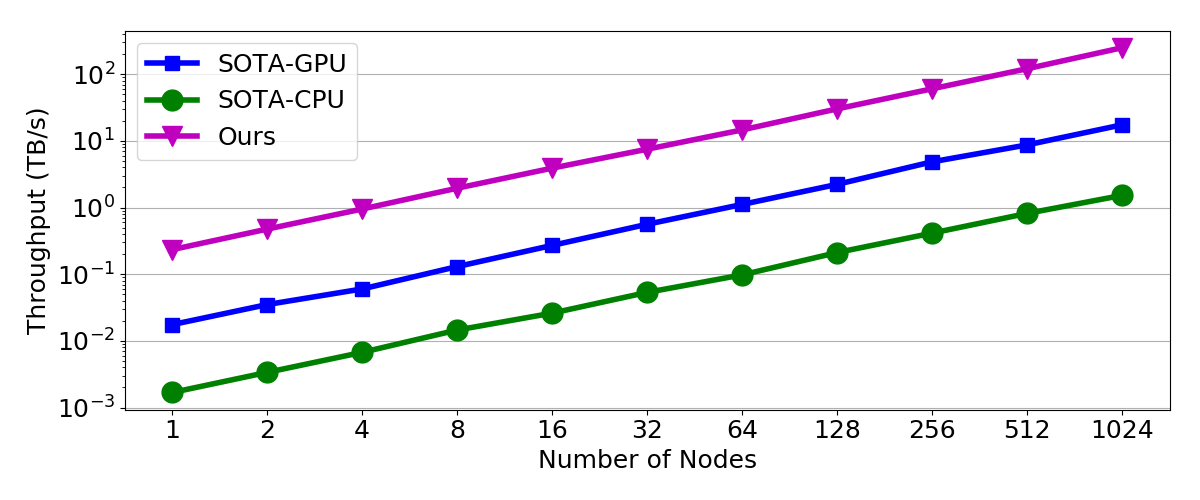}
    \caption{Aggregated data refactoring throughput at scale on Summit. 6 GPUs or 42 CPU cores are used per computing node, with each GPU or CPU core handling 1 GB in double precision.}
    \label{scale-smt}
\end{figure}

To show the potential of GPU-accelerated data refactoring in large-scale scientific applications, we conduct a weak scaling test on Summit. 
Here we parallelize the workload by assigning each GPU or CPU core an equal-sized data partition and perform decomposition and recomposition independently.
Due to the nature of multigrid-based data refactoring, parallelizing the workload in this way brings near linear speedups with negligible impact on decomposition and recomposition results. 
We assign each GPU or CPU core to one MPI process and perform data refactoring on 1 GB of simulation data.
For each computing node, we use the total available number of GPUs and CPU cores.
We scale the number of nodes up to 1024 in our tests on Summit.
As shown in Figure~\ref{scale-smt}, our optimized GPU data refactoring method achieves much greater throughput than state-of-the-art GPU and CPU designs. 
For example, we need only four computing nodes to achieve 1 TB/s data throughput, 
whereas state-of-the-art GPU and CPU designs require 64 and 512 nodes, respectively. 
With 1024 nodes (i.e., 6144 Volta GPUs), we achieve up to 250.26 TB/s aggregated data refactoring throughput.

\begin{figure*}[ht]
    \centering
    \begin{subfigure}[t]{0.47\textwidth}
    \includegraphics[width=\textwidth]{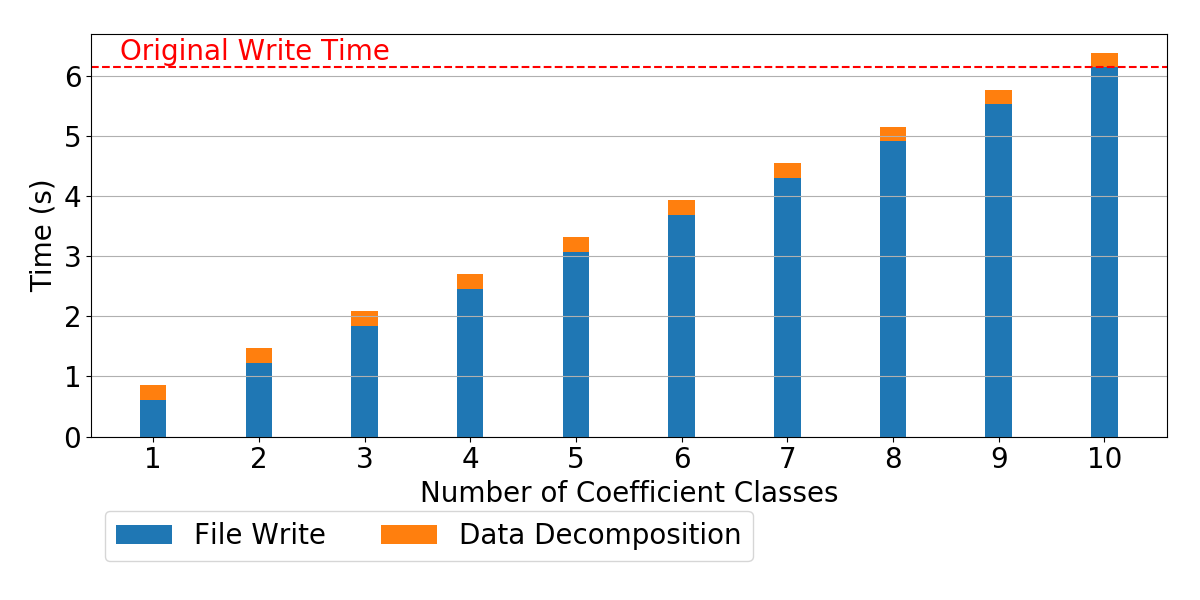}
    \vspace{-2em}
    \caption{Write simulation data}
    \end{subfigure}
    \begin{subfigure}[t]{0.47\textwidth}
    \includegraphics[width=\textwidth]{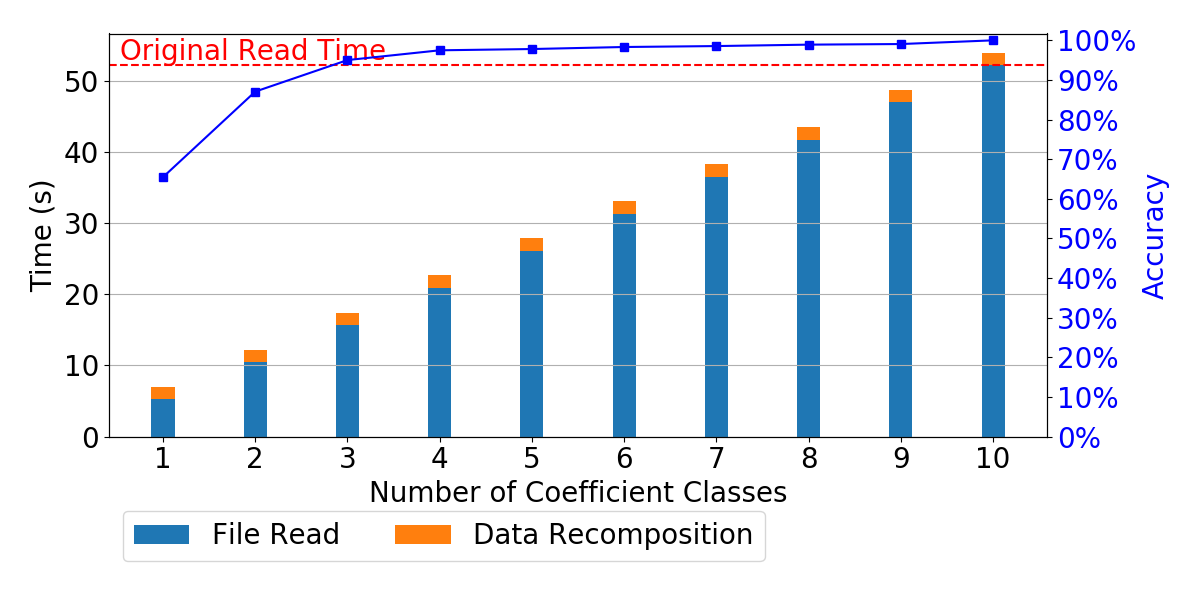}
    \vspace{-2em}
    \caption{Read simulation data and visualize}
    \end{subfigure}
    \caption{Showcase 1: Data refactoring in scientific visualization workflow}
    \label{vis-showcase}
    \vspace{-1em}
\end{figure*}

\begin{figure*}[ht]
    \centering
    \begin{subfigure}[t]{0.47\textwidth}
    \includegraphics[width=\textwidth]{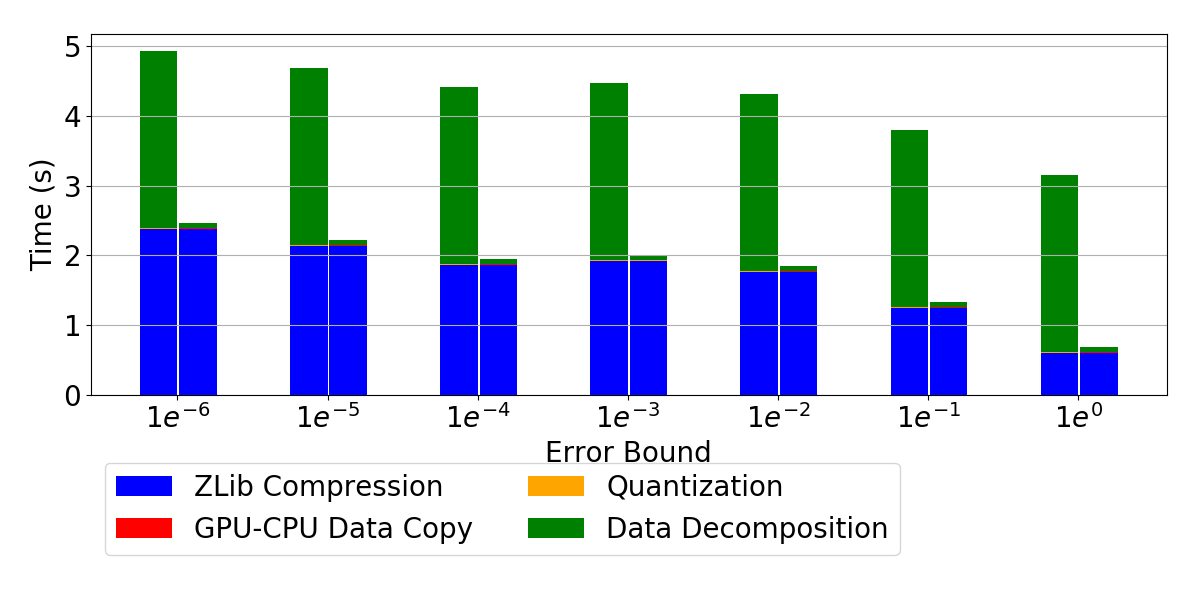}
    \vspace{-2em}
    \caption{Compression}
    \end{subfigure}
    \begin{subfigure}[t]{0.47\textwidth}
    \includegraphics[width=\textwidth]{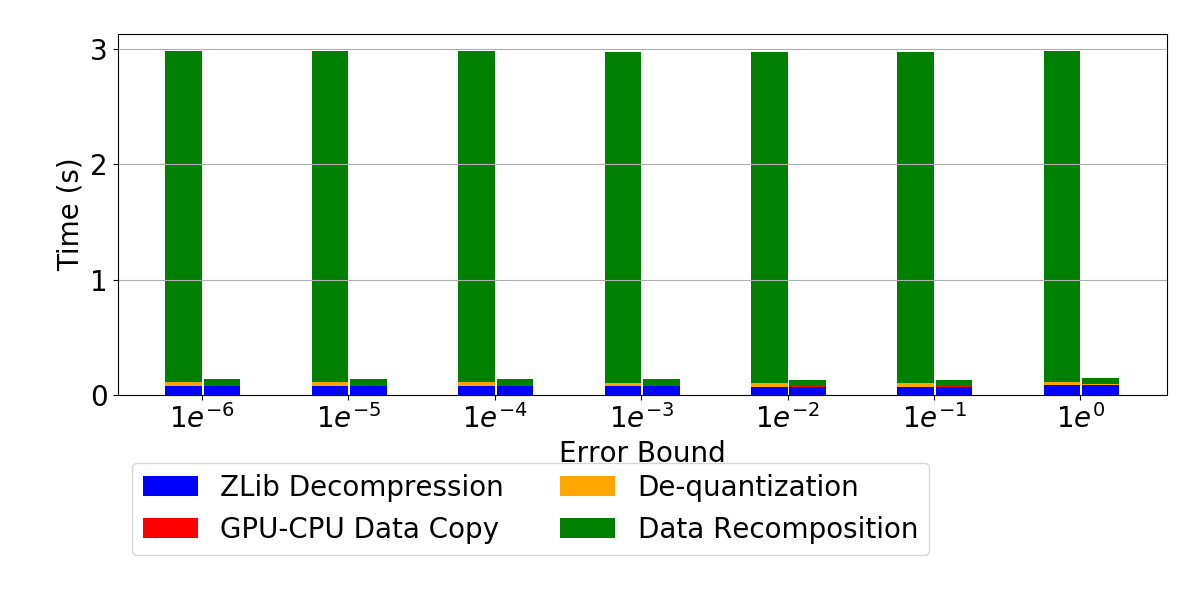}
    \vspace{-2em}
    \caption{Decompression}
    \end{subfigure}
    \caption{Showcase 2: MGARD lossy compression using CPU (left bars) vs.\ GPU (right bars)}
    \label{mgard-showcase}
    \vspace{-1.5em}
\end{figure*}

\section{Showcase}\label{sec:showcase}
Data refactoring algorithms were designed to offer much greater flexibility when managing large scientific data than the traditional methods.
With well-designed data management, data can be shared between scientific applications more intelligently with a large reduction in I/O costs.
However, inefficient data refactoring routines can diminish the benefits brought by data refactoring itself. 
Here we use two examples to show the benefits of GPU-based data refactoring over the CPU designs.

\subsection{Visualization workflow}
First we show how our GPU optimizations can make data refactoring effective when used for I/O cost reduction in scientific workflows that rely on file-based data sharing.
Figure~\ref{vis-showcase} shows the cost of writing and reading a 4 TB simulation data file using 4096 and 512 processes using the state-of-the-art ADIOS I/O library \cite{liu2014hello} on Summit with GPU-accelerated data refactoring enabled.
By writing or reading fewer coefficient classes, we can see immediate cost reduction in file write and read.
When our efficient GPU-accelerated data refactoring is used, we can see this reduction in the cost of file write and read can be effectively translated into a reduction in the total I/O cost.
Although multigrid-based data refactoring allows us to encode the most important information in the data with a few coefficient classes, it would not reduce the total I/O cost unless those coefficient classes can be efficiently computed or used for data recovery.
For example, in our experiments we achieve $\sim$95$\%$ accuracy for a chosen feature in the visualization result (i.e., the total area of the iso-surfaces~\cite{chen2019understanding, yakushin2020feature}) with only three out of ten coefficient classes.
This can be effectively translated into $\sim$66$\%$ I/O cost reduction.

\subsection{Lossy compression}
Multigrid-based hierarchical data refactoring can also be used as a preconditioner in scientific lossy compression software.
As one of the key components in lossy compression workflows, it is important to have efficient data refactoring in order to make fast lossy compression possible.
We showcase how our GPU-accelerated data refactoring can help improve the performance of lossy compression workflows in the MGARD lossy compression software.
MGARD is a CPU-based lossy compressor with three components in its workflow: multigrid-based data refactoring, quantization, and entropy encoding.
Figure~\ref{mgard-showcase} shows the time breakdown of the each component in MGARD \cite{mgard} when data refactoring remains on the CPU (left bars) or is off-loaded to the GPU (right bars).
In our test, besides the data refactoring process, we also off-load the quantization and de-quantization processes to the GPUs, since it can help reduce the GPU-CPU data transfer cost.
The entropy encoding stage (ZLib lossless compression) is kept on the CPU.
We can see that our GPU-accelerated data refactoring can greatly reduce the overall execution time of the lossy compression workflows.

\section{Related Work}

Multigrid-based data refactoring shares some similarities with multigrid solvers,
such as the use of multiple interlocking grids.
But while multigrid solvers aim to accelerate the solving of linear systems,
multigrid-based data refactoring aims to reconstruct scientific data progressively with hierarchical representations. 
This difference in focus leads to fundamental differences in both algorithms and optimization that prevent direct translation of
GPU optimizations. 

From an algorithmic perspective,
although data refactoring and multigrid solvers have some operations in common, data refactoring composes these operations in a unique way. 
Further, the correction used in data refactoring is designed specifically for the orthogonal projection, 
while the correction in multigrid solvers is used to generate the fine grid solution. From a GPU optimization perspective: optimizations for data refactoring need to consider handling large-volume scientific data, which means we need to consider not only limited GPU memory but also cases where refactoring process might share resources with original scientific computations on GPUs. So, it is essential to optimize for low memory footprint as well as performance.
Although part of the kernels used in data refactoring share similar computation patterns to those found in multigrid solvers, it is challenging to leverage existing work directly to achieve good parallelism and memory footprint balance in data refactoring.
For example, state-of-the-art GPU refactoring~\cite{mgard} uses a parallelization technique proposed by Basu et al.~\cite{Basu:2017}, which only use coarse grain vector-wise parallelism, which can cause lower performance for data refactoring. 
Although fine-grain parallelism has been achieved in previous works~\cite{Bell:2012,Esler:2012,Sebastian:2014,Richter:2015,Clark:2016}, it generally brings high memory footprint and it would require considerable effort to apply the optimizations to different algorithms.

\section{Conclusion}

We have presented optimized data refactoring kernels that allow for use of GPUs to
accelerate multigrid-based hierarchical refactoring for scientific data.
We evaluated our designs on two platforms, including the leadership-class Summit supercomputer at ORNL,
and showed that 
our GPU version can speed up data refactoring by up to 145$\times$ and 14$\times$ compared with state-of-the-art CPU and GPU designs, respectively,
and can achieve 250 TB/s throughput using 1024 nodes on Summit
We also showcased our work using a large-scale scientific visualization workflow and the MGARD lossy compression technique.  Together, these results demonstrate that scientists have another opportunity for dealing with their high data throughput requirements.  Inline refactoring of scientific data can offer performance improvements and temporal fidelity that can benefit a number of science scenarios.



\section*{Acknowledgment}
This work was made possible by support from the Department of Energy's Office of Advanced Scientific Computing Research, including via the CODAR and ADIOS Exascale Computing Project (ECP) projects.  This research used resources of the Oak Ridge Leadership Computing Facility, a DOE Office of Science User Facility supported under Contract DE-AC05-00OR22725.

\bibliographystyle{IEEEtran}
\bibliography{ref}

\end{document}